\documentclass[pra,twocolumn,a4paper,superscriptaddress,showpacs]{revtex4}
\usepackage{amsmath}
\usepackage{amsfonts}
\usepackage{color}
\usepackage{graphicx}

\newcommand{\bra}[2][]{#1\langle #2 #1\rvert}
\newcommand{\ket}[2][]{#1\lvert #2 #1\rangle}

\newcommand{\ketbra}[3][]{#1\lvert #2 #1\rangle #1\langle #3 #1\rvert}

\begin{document}
\title{Quantum methods for clock synchronization:  \\
Beating the standard quantum limit without entanglement}
\author{Mark \surname{de Burgh}}
\affiliation{School of Physical Sciences, The University of
Queensland, Queensland 4072, Australia}
\author{Stephen D. Bartlett}
\affiliation{School of Physics,
  The University of Sydney,
  New South Wales 2006, Australia}
\date{Received 17 June 2005; published 3 October 2005}

\begin{abstract}
  We introduce methods for clock synchronization that make use of
  the adiabatic exchange of nondegenerate two-level quantum systems:
  \emph{ticking qubits}.  Schemes involving the exchange of $N$ independent
  qubits with frequency $\omega$ give a synchronization
  accuracy that scales as $(\omega\sqrt{N})^{-1}$, i.e., as the standard
  quantum limit.  We introduce a protocol that makes use of $N_c$
  coherent exchanges of a single qubit at frequency $\omega$,
  leading to an accuracy that scales as
  $(\omega N_c)^{-1}\log N_c$.  This protocol beats
  the standard quantum limit without the use of entanglement, and we
  argue that this scaling is the fundamental limit for clock
  synchronization allowed by quantum mechanics.  We analyse the
  performance of these protocols when used with a lossy channel.
\end{abstract}

\maketitle

\section{Introduction}

Accurate clock synchronization is essential to a diverse range of
practical applications including navigation and global positioning,
distributed computation, telecommunications, and large science
projects like long baseline interferometry in radio astronomy.
Several recent works have explored the idea that concepts from
quantum information science may provide an advantage in clock
synchronization over ``classical''
approaches~\cite{Chu00,Joz00,Pre00,Gio01,Gio02,Yur02,Gio04a,Val04}.
A common idea in most of these schemes is to exploit entanglement in
some way to achieve an advantage, although some of these approaches
have been fraught with controversy as to the origin of their
advantage.

The aim of this paper is twofold.  First, we establish a framework
for exchanging quantum systems between parties who do not share
synchronized clocks, and demonstrate that clock synchronization
protocols based on such exchanges can be compared to a
\emph{standard quantum limit} (SQL), much like phase
estimation~\cite{Gio04}.  The SQL arises due to the statistics of
independent systems, and a standard assumption is that entanglement
(quantum correlations) between systems is required to beat the SQL.
Our second aim is to contradict this assumption: we introduce a
quantum method for clock synchronization that beats the SQL, yet
does not require the use of entanglement.  This result may allow for
practical implementations of these clock synchronization
protocols in the near future.

Suppose two parties, Alice and Bob, wish to synchronize their
clocks.  That is, each party has in their possession a
high-precision clock, both of which are assumed to run at exactly
the same \emph{rate} (frequency), but they do not agree on a common
time origin $t=0$.  Two classical methods for clock synchronization
are known as Einstein synchronization~\cite{Ein05} and Eddington
slow clock transport~\cite{Edd24}.  Our methods are based on the
latter, but for comparative purposes we first review Einstein
synchronization and its quantum developments.

\subsection{Einstein synchronization}

Einstein synchronization proceeds as follows. Alice records an
arbitrary time on her clock, $t_A$, and simultaneously sends a light
pulse to Bob, who records the time $t_B$ on his clock when he
received the pulse.  He reflects the pulse back to Alice, who
records the time on her clock when she receives it as $t'_A$. Alice
then sends to Bob the time $(t'_A + t_A)/2$, instructing him that it
is the time his clock should have been reading at time $t_B$. Thus
Bob obtains an estimate of $t_{BA}$, the time difference between
their clocks. This procedure is repeated many times and the results
averaged to obtain an accurate estimate of $t_{BA}$.  A modern
technique using Einstein synchronization is known as Time Transfer
by Laser Link~\cite{Sam98}, which predicts that ground stations
communicating their light pulses to a common satellite can
synchronize within 100~ps. A substantial part of the uncertainty in
this method is due to short term variation in message delivery time
as a result of atmospheric changes in the refractive index.

If many independent uses of such a protocol are averaged, the
accuracy is determined by the central limit theorem.  For a Gaussian
shaped coherent state laser pulse with frequency spread $\Delta
\omega$, and an average over the arrival times of $N$ independent
photons in the laser pulse, the uncertainty in time synchronization
for large $N$ is~\cite{Gio04}
\begin{equation}
  \Delta t = \frac{1}{\Delta \omega \sqrt{N}}\,.
\end{equation}
This scaling of $1/\sqrt{N}$ is commonly known as the \emph{standard
quantum limit} (SQL), also known as the ``shot noise'' limit when
referring to optics.  As we will show, the SQL can be expressed
another way: to obtain time synchronization to $k$ bits of precision
requires transmission of $O(2^{2k})$ photons for a fixed frequency
spread~\footnote{The Big $O$ notation indicates that the function
gives an upper bound asymptotically. Explicitly, $f(x)$ is $O(g(x))$
if there are constants $c$ and $x_0$, such that for all $x
> x_0$, $f(x) < cg(x)$.}.  It is generally accepted that
``classical'' (i.e., independent, unentangled) strategies cannot
beat the SQL~\cite{Gio01}.

Recently the use of concepts from quantum information~\cite{Nie00},
in particular the use of entangled states of quantum systems, has
been revolutionizing the theoretical limits of precision
measurements, and clock synchronization is no exception.  Instead of
classical coherent state light pulses, one can use highly entangled
states of many photons and beat the SQL; see~\cite{Gio01}.
Essentially, the advantage is due to entanglement-induced bunching
in arrival time of individual photons, enabling more accurate timing
measurements. The key disadvantage of this technique is that the
loss of a single photon destroys the entanglement and renders the
measurement useless~\cite{Gio04,Gio01} (although techniques have
been developed to ``trade off'' the quantum advantage in return for
robustness against loss~\cite{Gio02}). Furthermore, the effect of
dispersion is known to be an important issue with quantum-enhanced
Einstein protocols, with the use of entanglement possibly offering
an advantage here as well~\cite{Fit02,Gio04b}.  We note that
frequency entanglement across large numbers of photons is
experimentally challenging.  Thus, it is worthwhile to consider
alternate methods, such as those based on Eddington's protocol.

\subsection{Eddington slow clock transport}

The second traditional method for clock synchronization is known as
the Eddington slow clock transport. In this protocol, Alice
synchronizes a ``wristwatch'' (a clock that can easily be
transported) with her own clock, and then
adiabatically~\footnote{i.e., the Hamiltonian of the ticking qubit
must be constant throughout the protocol.} transports the wristwatch
to Bob. Bob can then determine the time difference between his clock
and the wristwatch and hence obtain an estimate of $t_{BA}$.  The
principal advantage of this method over Einstein synchronization is
that the accuracy is inherently independent of the message delivery
time.

Recently, Chuang~\cite{Chu00} proposed two protocols which are
quantum versions of the Eddington slow clock transport.  In these
protocols, the wristwatch is realized by \emph{ticking qubits:}
nondegenerate two-level quantum systems that undergo time evolution.
The first protocol of~\cite{Chu00} requires $O(2^{2k})$ ticking
qubit communications (a coherent transfer of a single qubit from
Alice to Bob) to achieve an accuracy in $t_{BA}$ of $k$ bits. The
protocol requires no entangled operations or collective
measurements, with synchronization accuracy that scales as the SQL.
The second protocol presented in~\cite{Chu00} makes use of the
Quantum Fourier Transform~\cite{Nie00} and an exponentially large
range of qubit ticking frequencies.  This protocol requires only
$O(k)$ quantum messages to achieve $k$ bits of precision, an
exponential advantage over the SQL.  Although these schemes give
insight into the ways that quantum resources may allow an advantage
in clock synchronization, they are unsatisfactory for two reasons:
(1) the first scheme does not beat the SQL, while the second
scheme's use of exponentially demanding physical resources is
arguably the origin of the enhanced efficiency~\cite{Chu00,Gio01};
and (2) Alice and Bob need to \emph{a priori} share a synchronized
clock in order to implement the required operations. We will show
how both of these problems can be overcome.

In this paper we perform an extensive analysis of the use of ticking
qubits in clock synchronization.  We express all operations on the
ticking qubits in a rotating frame, reducing the problem of clock
synchronization to one of phase estimation. Thus, we build on the
wealth of knowledge which has been developed around phase estimation
in interferometry~\cite{Cav81,Yur86,Hol93,San95,Bol96,Ber00} and
also on the establishment of a shared reference
frame~\cite{Per01,Bag01,Rud03}. These techniques essentially
determine optimal entangled input states and collective measurements
and can directly yield corresponding clock synchronization
algorithms. We also note that there is a direct connection to Ramsey
interferometry, where it has been shown~\cite{Bol96} that maximally
entangled states of $N$ two-level quantum systems yield a frequency
estimate that beats the SQL.

The entanglement required to gain such an advantage has been
recently demonstrated between three~\cite{Mit04,Lei04} and
four~\cite{Sac00,Wal04} qubits, but the difficulty of producing
complex entangled states and collective measurements for large
numbers of qubits currently limits the practical use of these
protocols.  However, recent work in techniques for reference frame
alignment have demonstrated that comparable advantages can be gained
\emph{without} the need for highly entangled states or collective
measurements, provided that coherent two-way communication is
allowed~\cite{Rud03}.  We present a protocol that, through the use
of coherent communications of a single qubit, beats the SQL without
the use of entanglement.

\subsection{Assumptions and Conventions}

Throughout this paper, we will assume that Alice and Bob share an
inertial reference frame and thus relativistic effects are ignored.
We will also assume that Alice and Bob's clocks are classical in the
sense that they are not appreciably affected by their use in state
preparations and measurements.  Practically, one may consider
Alice's and Bob's clocks to be realized by large-amplitude lasers at
a common frequency~\cite{Wis03,Wis04}.

For concreteness, our ticking qubits are realized by two electronic
energy levels of an atom (i.e., a standard two-level atomic qubit
such as those described in, for example, the proposal for an
ion-trap quantum computer by Cirac and Zoller~\cite{Cir95}), as
these most simply illustrate our discussion. However, all that is
required of a ticking qubit is that it have, as a basis, two
non-degenerate energy levels and a Hamiltonian that can be
considered to be constant throughout the protocol.   Perhaps the
most useful implementation of the ticking qubit would be an optical
qubit represented by the presence or absence of a single photon in a
given propagating mode.

We adopt the following conventions common to the quantum information
community~\cite{Nie00}.  The two energy eigenstates of a qubit are
labeled by the computational basis states $\ket{0}$ and $\ket{1}$,
with their energy eigenvalues  assumed to be such that $E_1 < E_0$.
(Note, with this convention, $\ket{0}$ is the excited state.)  We
define the Pauli $Z$ operator as $Z \ket{0} = \ket{0}$ and $Z
\ket{1} = -\ket{1}$.

The Hamiltonian for our ticking qubits is $H_0=\hbar \omega Z / 2$.
The evolution is described by the Schr\"odinger equation $i \hbar
\frac{d}{dt}\ket{\psi} = H_0 \ket{\psi}$.  If a qubit is initiated
in the state $\ket{\psi} = \frac{1}{\sqrt{2}}[\ket{0} + \ket{1}]$,
then one can picture the Bloch vector rotating anti-clockwise about
the $z$-axis with an angular frequency of $\omega$.  For
convenience, we choose to work in a rotating frame (interaction
picture), in which states are described as $\ket{\psi}_I = e^{i H_0
t / \hbar} \ket{\psi}$, and observables and transformations as $A_I
= e^{iH_0 t / \hbar} A e^{-iH_0 t / \hbar}$.  (In what follows, we
will drop the subscript $I$, as we will be working exclusively in
the interaction picture.) In this rotating frame, our qubits no
longer tick, and the problem of clock synchronization is reduced to
one of phase estimation.

Because we are using qubits of a fixed frequency $\omega$ to perform
clock synchronization, and $\phi = \omega t_{BA}$ can only take
values between 0 and $2\pi$, it is clear that we can only
synchronize within an interval $0 \leq t_{BA} \leq 2\pi / \omega$.
We will assume that Alice and Bob's clocks are already synchronized
to within one half a period ($0 \leq t_{BA} \leq \pi/\omega$) and
the goal is to synchronize them more accurately.

\section{Framework for clock synchronization using ticking qubits}

When describing protocols between two parties with unsynchronized
clocks, one must take care in expressing how state preparations and
operations performed by one party should be represented by the
other~\cite{BRS03}.  In particular, if a quantum operation performed
by Alice is defined relative to her classical clock (often not
explicitly mentioned in other works), then this quantum operation
will be expressed differently by another observer, Bob, whose clock
is not synchronized with Alice's.  We now derive the transformation
laws between parties with different clocks for the case of two-level
atomic qubits, and we will explicitly see the role played by the
phase of the classical clock (laser) in performing state
preparations, operations and measurements.  For further details and
a different perspective, see~\cite{Enk05}.

\subsection{Operations using Rabi pulses}

In order to define operationally how a ticking qubit is
correlated with a classical clock, we now describe in detail
how preparations, operations and measurements are done on this
system using a laser (the classical clock, considered a part of the
experimental apparatus).  Single qubit operations are performed by
tuning a laser to the $\ket{0}\rightarrow \ket{1}$ transition,
introducing Rabi flopping between the states at the Rabi frequency
$\Omega$~\cite{Cir04}. We assume an interaction picture Hamiltonian
of,
\begin{equation}
  H_{\rm Rabi}(\phi) = \frac{\hbar\Omega}{2}\Bigl[e^{-i\phi}\ketbra{0}{1} +
  e^{+i\phi}\ketbra{1}{0}\Bigr]\,,
\end{equation}
where $\phi$ is the phase angle of the laser, defined relative to
that party's clock and which can be varied as part of the
experimental apparatus.  Because Alice and Bob do not share
synchronized clocks, their phase references will be different, and
the process of clock synchronization will amount to determining the
difference between these phase references. To be notationally clear,
we write $\phi_P$ to represent an angle relative to party $P$'s
phase reference.  Then $\phi_{BA} = \omega t_{BA}$ is the difference
between Bob and Alice's phase references, i.e., between what Alice
defines to be $\phi_A = 0$ and what Bob defines to be $\phi_B=0$.

If the laser with phase $\phi_P$ interacts with the atom for a time
$t$, the effective
unitary evolution is $U(t,\phi_P) = \exp(-iH_{\rm Rabi}(\phi_P)t /
\hbar)$. A laser pulse maintained for time $t = k\pi /\Omega$ is
known as a ($k\pi$)-pulse, with a unitary operator $\Pi_k (\phi_P)$
given by,
\begin{equation}
 \Pi_k (\phi_P) = \begin{pmatrix}
    \cos(k\pi/2) & -ie^{-i\phi_P}\sin(k\pi/2) \\
    -ie^{+i\phi_P}\sin(k\pi/2) & \cos(k\pi/2)
    \end{pmatrix}\,,
\end{equation}
expressed in the computational (energy eigenstate) basis.  Consider
the unitary transformation matrix for a $\pi$-pulse ($k=1$):
\begin{equation}
    \Pi_1 (\phi_P) = \begin{pmatrix} 0 & -ie^{-i\phi_P} \\
    -ie^{+i\phi_P} & 0 \end{pmatrix} \,,
\end{equation}
which, up to an overall phase factor, gives,
\begin{equation}
  \label{Eqn:XOperation}
  \Pi_1(\phi_P) = e^{-i \phi_P Z/2}
  \begin{pmatrix} 0 & 1 \\ 1 & 0 \end{pmatrix} e^{+i \phi_P Z/2} \,.
\end{equation}
We define the operation $\Pi_1(\phi_P = 0)$ to be the Pauli $X$
operation for the party $P$, denoted $X_P$.  Note that this
operation depends on the phase $\phi_P$ of party $P$. (Specifically,
it depends on the phase of the classical laser pulse used to perform
the operation.)  Thus, the Pauli $X$ operator for party $P$ is
defined \emph{relative} to $P$'s clock, and, in general, different
parties with unsynchronized clocks will define such operators
differently. We contrast this result with the Pauli $Z$ operator,
which is diagonal in the energy basis and is defined independently
of any clock.

We will also make use of the $\pi/2$-pulse ($k=1/2$),
\begin{align}
 \Pi_{(1/2)} (\phi_P) &= \frac{1}{\sqrt{2}} \begin{pmatrix}
    1 & -ie^{-i\phi_P} \\
    -ie^{+i \phi_P } & 1 \end{pmatrix} \nonumber  \\
    &= e^{-i\phi_P Z/2} \frac{1}{\sqrt{2}}\begin{pmatrix}
    1 & -i \\
    -i & 1 \end{pmatrix} e^{+i \phi_P Z / 2} \,,
\end{align}
and specifically the operation
\begin{equation}
    H_P \equiv \Pi_{(1/2)} (\phi_P=\pi/2) = \frac{1}{\sqrt{2}}
    \begin{pmatrix}
    1 & -1 \\
    1 & 1
    \end{pmatrix} \,.
\end{equation}
This operation is essentially a Hadamard gate~\cite{Nie00}. Again,
we note that this operation $H_P$ is defined relative to the party
$P$'s clock.

In general, we use the notation $U_P$ to denote an operation
performed by an appropriate pulse, where the phase angle of the
laser pulse was measured with respect to party $P$'s clock. Because
Alice and Bob do not share synchronized clocks, in general $U_A \neq
U_B$.

\subsection{Defining states and projective measurements}

Because operations are defined relative to the clock (laser) used to
perform them, quantum states will also depend on this reference.
Here, we will demonstrate how states (and the Bloch sphere) are
defined relative to party $P$'s clock.

First we observe that the energy eigenstates $|0\rangle$ and
$|1\rangle$ are defined the same for any party, independent of their
clocks.  Also, from the previous section we have a well defined
notion of an operation $U_P$ performed by party $P$.  Thus, each
party can define a general state on the Bloch sphere in terms of the
operation $U_P$ needed to create that state from the eigenstate
$\ket{0}$. For example party $P$ will define the state
$\ket{\psi_0}_P = \frac{1}{\sqrt{2}}[\ket{0} + \ket{1}]_P$ as the
state produced by performing their $H_P$ operation on the state
$\ket{0}$. In general, two parties will differ in how they describe
a given state, because each will describe it relative to their own
clock.

Projective measurements on a single qubit are defined similarly.  A
projective measurement in an arbitrary basis $\{ |\psi\rangle_P,
|\psi^\perp\rangle_P \}$ for a party $P$ can be viewed as follows:
first, they perform the operation $U_P$ which transforms this basis
to the computational basis, and then measure in this basis.  (Note
that this procedure is precisely how projective measurements are
performed in~\cite{Cir95}.)  Such a definition is consistent with
the definition of states given above.  For example, the state
$\ket{\psi_0}_P = \frac{1}{\sqrt{2}}[\ket{0} + \ket{1}]_P$ can also
be defined as the unique state such that, if a party $P$ performs
the operation $H_P$ and then measures in the computational basis,
they obtain the result $|1\rangle$ with certainty.

\subsection{Bipartite operations without synchronized clocks}

Suppose that Bob's clock differs from Alice's by an amount $t_{BA}$.
If $\phi_{BA} = \omega t_{BA} \neq 0$, Alice and Bob do not describe
general states and operations equivalently.  For example, if Alice
performs her $H_A$ operation on $\ket{0}$, passes the state to Bob,
and Bob performs his $H_B$ operation on this state and measured the
result in the computational basis, he will not get the state
$\ket{1}$ with certainty.

Their operations are, however, related by
\begin{align}
  U_B(\phi_B) &= U_A(\phi_A + \phi_{BA}) \nonumber \\
  &= e^{-i \phi_{BA} Z /2} U_A(\phi_A) e^{+i \phi_{BA} Z / 2} \,.
  \label{Eqn:BobsOpAsAlicesOp}
\end{align}
We can interpret this result as follows:  for Alice to perform the
same operation as Bob, she ``devolves'' backwards in time by an
amount $t_{BA}$, performs the same operation with her laser (i.e.,
relative to her phase reference), and then evolves forward in time
by an amount $t_{BA}$.  With this transformation rule between
operators, it is straightforward to show that the state
$|\psi\rangle_B$ that Bob assigns to a system is related to the
state $|\psi\rangle_A$ assigned by Alice according to
\begin{equation}
  \label{TransState}
  \ket{\psi}_B = e^{-i \phi_{BA} Z /2} \ket{\psi}_A\,.
\end{equation}

\section{Clock Synchronization Protocols}

In this section, we present a number of different clock
synchronization protocols based on the exchange of ticking qubits,
and compare their performance and resource requirements.  First, we
present a simple protocol with uncertainty that achieves the SQL,
followed by a slight modification with similar performance that will
be useful for comparative purposes.  We then introduce an improved
protocol that beats the SQL, and yet does not require entanglement.

\subsection{Simple One-way Protocol}
\label{sec:Simple}

The simplest ticking qubit clock synchronization protocol based on
Eddington's slow clock transport proceeds as follows.  Alice
prepares a ticking qubit in the energy eigenstate $\ket{0}$ and
performs her operation $H_A$, producing the state $\ket{\psi}_A =
\frac{1}{\sqrt{2}}[\ket{0} + \ket{1}]_A$.  The qubit begins to
``tick,'' i.e., evolve under the Hamiltonian $H_0 = \hbar\omega
Z/2$, but we do not express this evolution as we are working in the
interaction picture.

Alice then sends the qubit to Bob, who represents this state
(using Eq.~(\ref{TransState})) as
\begin{align}
  \ket{\psi}_B &= e^{-i \omega t_{BA} Z /2} \ket{\psi}_A \\
  &= \tfrac{1}{\sqrt{2}} \bigl[e^{-i \omega t_{BA}/2}\ket{0} + e^{+i \omega
  t_{BA}/2}\ket{1}\bigr]_B \,.
\end{align}
Bob performs the operation $H_B$, yielding
\begin{equation}
  \ket{\psi}_B = \bigl[ -i \sin (\omega t_{BA} / 2 )\ket{0} +
  \cos (\omega t_{BA}/2)\ket{1} \bigr]_B \,.
\end{equation}
Bob then measures the observable $O_B = -Z$. The expected value of
this observable is:
\begin{equation}
  \label{eqn:ExpectOSimpleOneWay}
  \langle O_B \rangle = {}_B\bra{\psi_1} O_B \ket{\psi_1}_B
  = \cos{(\omega t_{BA})} \,.
\end{equation}
The uncertainty in the observable is
\begin{equation}
  \Delta O_B = \sqrt{\langle O_B^2 \rangle -
  \langle O_B \rangle ^2} = \sin{(\omega t_{BA})}\,.
\end{equation}

An estimate of $t_{BA}$ is obtained from an estimate of the phase
angle $\omega t_{BA}$ in (\ref{eqn:ExpectOSimpleOneWay}).  To
unambiguously determine a value for $t_{BA}$, we require an initial
time synchronization accurate to within the range $0 \leq t_{BA}
\leq \pi / \omega$. Practically, it is useful to restrict $t_{BA}$
further, such as in the range $\pi/6 \leq \omega
t_{BA} \leq 5\pi /6$, i.e., to ``lie on the fringe.'' This method
leads to an uncertainty in his estimate of $t_{BA}$ of $\Delta
t_{BA} = 2 /\omega$.

The procedure is repeated $N$ times and the results averaged.
Because each measurement may be considered an independent random
variable, the central limit theorem tells us that the uncertainty
scales as $1/\sqrt{N}$, giving a final uncertainty after $N$
iterations of $\Delta t_{BA} = 2/(\omega \sqrt{N})$. It will be
useful for later comparisons to count the number of \emph{qubit
communications} rather than the number of iterations, as is standard
in analyses of quantum communication complexity.  In this case, the
number of iterations $N$ is equal to the number of qubit
communications $N_c$ and
\begin{equation}
  \label{eqn:DtSimpleOneWay}
  \Delta t_{BA} = \frac{2}{\omega \sqrt{N_c}}\,.
\end{equation}
The protocol relies purely on classical statistics and scales the
same as the straightforward Einstein protocol, i.e., at the SQL.
However, because it is an Eddington-type protocol, it may perform
better in some situations, i.e., when there is large message
delivery time uncertainty.

\subsection{Bits of precision}

It will be useful to quantify the performance of this protocol in an
alternate way to Eq.~(\ref{eqn:DtSimpleOneWay}).  We will determine
the resources required to determine the phase $\phi_{BA} = \omega
t_{BA}$ to $k$ bits of precisions with some probability of error. In
the above simple protocol, let $P_1$ be the probability that Bob
measures the ticking qubit the state $\ket{1}$. The Chernoff
bound~\cite{Nie00} tells us that, after $N$ independent iterations,
the probability that the difference between Bob's estimate
$\overline{P}_1$ and the true value $P_1$ is greater than some
precision $\delta$ decreases exponentially with $N$, specifically,
\begin{equation}
  \label{eq:Chernoff1}
  Pr[|\overline{P}_1 - P_1| \ge \delta]\le 2e^{-N\delta^2/2}\,.
\end{equation}
The observable $O_B$ that Bob measures is related to this
probability by $\langle O_B \rangle = 2P_1 - 1$. Thus, his estimate
$\overline{\langle O_B \rangle}$ is related to his probability
estimate by $\overline{\langle O_B \rangle} = 2\overline{P_1}-1$.
Thus, $|\overline{\langle O_B \rangle} - \langle O_B \rangle | =
2|\overline{P_1} - P_1|$, and
\begin{equation}
  \label{eq:ChernoffO}
  Pr\bigl[\bigl|\overline{\langle O_B \rangle} - \langle O_B \rangle \bigr| \ge
  2\delta\bigr] \le 2e^{-N\delta^2/2}\,.
\end{equation}

We require this bound on the observable $O_B$ to give a bound on the
accuracy of our estimate of $t_{BA}$.  As above, we require an
initial synchronization within the range $\pi/6 \leq \omega t_{BA}
\leq 5\pi /6$.  Thus, if $|\overline{\langle O_A \rangle} - \langle
O_A \rangle| < 2\delta $, then $|\overline{t_{BA}} - t_{BA}| <
4\delta / \omega$, and Eq.~(\ref{eq:ChernoffO}) gives
\begin{equation}
 \label{eq:Chernoff2}
  Pr[|\overline{t_{BA}} - t_{BA}| \ge 4\delta/\omega]\le 2e^{-N \delta^2/2}\,.
\end{equation}
Then the number of iterations $N$ required to estimate $t_{BA}$ with
precision $4\delta/\omega$ with probability of error bounded by
$\epsilon$ is given by,
\begin{equation}
  N = \frac{2 \ln (2 / \epsilon)}{\delta^{2}} \,.
\end{equation}
Let $T = \phi_{BA}/\pi = \omega t_{BA}/\pi$. If we require an
estimate of $T$ to $k$ bits of precision, then $4\delta/\pi =
2^{-k}$, and expressing in terms of the number of qubit
communications $N_c$ required gives
\begin{equation} \label{eq:SimpleAlgorithmCost}
  N_c(k) = \frac{32}{\pi^2} \ln (2 / \epsilon)2^{2k} \,.
\end{equation}
This result can be considered as a reexpression of the SQL.  To
achieve $k$ bits of precision in an estimate, the SQL states that
one needs $O(2^{2k})$ iterations of the protocol.

\subsection{Simple Two-way Protocol}
\label{sec:Simple2}

It will be useful to modify the ``one-way'' protocol defined above
into a two-way protocol, as follows.  As before, Alice prepares her
ticking qubit in the energy eigenstate $\ket{0}$ and performs her
operation $H_A$ before sending the qubit to Bob.  Rather than
measuring this qubit, Bob performs his operation $X_B$ and sends the
qubit back to Alice.  She then performs her operation $X_A$.  The
resulting combined transformation $X_A X_B$ is described in Alice's
frame as
\begin{align}
  X_A X_B &= X_A (e^{-i\omega t_{BA} Z/2} X_A e^{+i\omega t_{BA}Z/2})
  \nonumber \\
  &= e^{+i \omega t_{BA} Z} \,.
\end{align}
(This joint operation will be the key basic component of our
improved protocol.)  Finally, Alice performs her $H_A$ operation and
measures the observable $O_A = -Z$.  Here, we require the initial
uncertainty in $t_{AB}$ to be half of that described in the protocol
of Sec.~\ref{sec:Simple}.  The expected value of this observable is
$\langle O_A \rangle = \cos(2\omega t_{BA})$, yielding an
uncertainty in $t_{BA}$ of $\Delta t_{BA} = 1 / \omega$. Thus, after
performing this two-way operation $N$ times and averaging the
results, we have $\Delta t_{BA} = 1/(\omega \sqrt{N})$.  Expressed
in terms of the number of qubit communications $N_c$, we have
\begin{equation}
  \label{eqn:DtSimpleTwoWay}
  \Delta t_{BA} = \frac{\sqrt{2}}{\omega \sqrt{N_c}} \,.
\end{equation}

We can also analyse this protocol in terms of number of qubit
communications required to achieve a time synchronization of $k$
bits with probability of failure $\epsilon$, yielding
\begin{equation} \label{eq:SimpleAlgorithmCost2}
  N_c(k) = \frac{16}{\pi^2} \ln (2 / \epsilon)2^{2k}\,.
\end{equation}
This protocol, too, scales as the SQL due to the fact that the
iterations of the protocol are independent.

We note that this two-way procedure is very similar to the ``Ticking
Qubit Handshake'' (TQH) protocol of Chuang~\cite{Chu00}. However,
in~\cite{Chu00}, the two local ``clock-dependent'' transformations
$X_A$ and $X_B$ are described in the same frame, and thus Alice and
Bob require \emph{a priori} synchronized clocks to implement the
described operations.  Our procedure is expressed entirely in terms
of operations by parties who do not share synchronized clocks, and
demonstrates that the required operations can indeed be performed
without prior synchronized clocks.

\subsection{Improved Clock Synchronization Protocol}
\label{sec:Improved}

We now present an improved clock synchronization protocol that beats
the SQL.  A standard assumption is that entanglement between qubits
is required in order to beat this limit.  However, the protocol we
introduce does not require entanglement, instead relying on an
increased complexity in coherent communications. This protocol is
based on the work of~\cite{Rud03}, which investigated the related
problem of quantifying the resource requirements for establishing a
shared Cartesian frame.

In this protocol, Alice and Bob use a phase estimation algorithm that
estimates each bit of the phase angle independently. We define the
phase angle $\omega t_{BA} = \pi T$, where $T$ has the binary
expansion $T = 0.t_1t_2t_3\cdots$.  Alice and Bob will attempt to
determine $T$ to $k$ bits of precision, and accept a total error
probability $P_{\rm error} \leq \epsilon$.  If the total error
probability is to be bounded by $\epsilon$, then each $t_i$,
$i=1,\ldots,k$, must be estimated with an error probability of
$\epsilon /k$. (An error in any one bit causes the protocol to fail,
so the total error probability in estimating all $k$ bits is $P_{\rm
error} = 1-(1-\epsilon /k)^k \le \epsilon$.)

To estimate the first bit $t_1$, Alice and Bob use the two-way
protocol defined in Sec.~\ref{sec:Simple2}.  Expressing
$\langle O_A\rangle$ in terms of $T$, we have
\begin{equation}
  \langle O_A \rangle = \cos{(2 \omega t_{BA})}
  = \cos{(2\pi 0.t_1 t_2\cdots)}
\end{equation}
By sending $n_1$ qubits and averaging the results, Alice obtains
$\overline{\langle O_A \rangle}$, the estimate of $\langle O_A
\rangle$. If $n_1$ is chosen such that $|\overline{\langle {O}_A
\rangle} - \langle O_A \rangle| \le 1/2$ with some error
probability, then $|\overline{T} - T  | \le 1/4$, determining the
first bit $t_1$ with this same probability. The required number of
iterations $n_1$ is given by the Chernoff bound
(\ref{eq:ChernoffO}), with $\delta = 1/4$,
\begin{equation}
    \label{FirstBitError}
    Pr\Bigl[|\overline{\langle O_A \rangle} - \langle O_A \rangle|
    \ge 1/2\Bigr] \le \epsilon/k \le 2e^{-n_1/32} \,,
\end{equation}
giving $n_1 \ge 32\ln(2k/\epsilon)$.

Now we define a similar procedure for estimating an arbitrary bit,
$t_{j+1}$. Alice prepares the energy eigenstate $\ket{0}$, and
performs her $H_A$ operation. Alice and Bob then pass the qubit back
and forth to each other $2^j$ times, each time Bob performs his
$X_B$ operation and Alice performs her $X_A$ operation. That is,
they jointly implement the operation $(X_A X_B)^{2^j}$.  Finally
Alice performs her $H_A$ operation. Expressing these operations in
Alice's frame, the protocol to estimate $t_{j+1}$ produces the state
\begin{align}
 \ket{\psi_j}_A &= H_A (X_A X_B)^{2^j} H_A \ket{0} \nonumber \\
 &= H_A (e^{+i\omega t_{AB} Z})^{2^j} H_A \ket{0} \nonumber \\
 &= H_A e^{+i2^j \omega t_{BA} Z} H_A \ket{0} \nonumber \\
   &= \bigl[ i\sin (2^j \omega t_{BA})\ket{0} +
  \cos (2^j \omega t_{BA} )\ket{1}\bigr]_A \,.
\end{align}
Alice then measures the observable $O_A = -Z$. The expected value of
this observable is:
\begin{align}
     \langle O_A \rangle
     &= \cos{(2^{j+1} \omega t_{BA})} \nonumber \\
     &= \cos(2^j[2\pi 0.t_1t_2\cdots] ) \nonumber \\
     &= \cos(2\pi t_1t_2\cdots t_j.t_{j+1}t_{j+2}\cdots ) \nonumber \\
     &= \cos(2\pi 0.t_{j+1}t_{j+2}\cdots ) \,. \label{Eqn:ExpectOImproved}
\end{align}
This expression has the same form as one iteration of the scheme to
estimate the first bit $t_1$; Alice and Bob simply require more
exchanges to implement $(X_A X_B)^{2^j}$.  To get a probability
estimate for each bit $t_{j+1}$, this more complicated procedure is
repeated $n_{j+1}$ times.  Because we require equal probabilities
for correctly estimating each bit, we can set all $n_{j+1}$ equal to
the same value, $n \sim 32 \ln(2k/\epsilon)$.

The total number of qubit communications required to estimate $T$ to
$k$ bits of precision with total error probability less than
$\epsilon$ is thus
\begin{align}
  \label{eq:NQubitCommunications}
  N_c &= 2n \sum_{j=1}^k 2^{j-1} = 2n(2^k - 1) \nonumber \\
  &= 64\ln(2k/\epsilon)(2^k - 1) \,,
\end{align}
which scales as $O(2^{k} \ln (2k/\epsilon))$.  We note that, unlike
in the previous protocols, the majority of these qubit
communications (i.e., those used to estimate each bit $j$) must be
performed coherently using the \emph{same} qubit.

It is useful to connect this result to the uncertainty in the time
synchronization $\Delta t_{BA}$. We model our process as giving a
successful estimate, $t_{\rm est}$ (i.e., $T$ within $k$ bits of
precision, $\Delta t_{\rm est} \leq 2^{-k} \pi / \omega$) with
probability $(1-\epsilon)$ and a random time estimate $t_{\rm rand}$
(within one half period of our qubit ticking frequency) with a
probability $\epsilon$ giving $t = (1-\epsilon) t_{\rm est} +
\epsilon t_{\rm rand}$. If we assume the errors in $t_{\rm est}$ and
$t_{\rm rand}$ are independent then
\begin{align}
  \Delta t_{BA} &= \sqrt{(1-\epsilon)^2 (\Delta t_{\rm est})^2 + \epsilon^2
  (\Delta t_{\rm rand})^2} \nonumber \\
  &= \sqrt{(1-\epsilon)^2 2^{-2k} \pi^2 / \omega^2 +
  \epsilon^2  \pi^2 / \omega^2} \,.
  \label{eq:GeneralUncertaintyEquation}
\end{align}
We must now choose an $\epsilon$ for each $k$, so that under the
constraints (\ref{eq:GeneralUncertaintyEquation}) and
(\ref{eq:NQubitCommunications}), $\Delta t_{BA}$ decreases inversely
with the largest possible function of $N_c$.  If we choose $\epsilon
= 1 / 2^k$ then $\omega\Delta t_{BA} = O(2^{-k})$. Using
(\ref{eq:NQubitCommunications}) then gives $N_c = k 2^k$, and
ignoring terms logarithmic in $k$, gives an uncertainty which scales
as
\begin{equation}
  \label{eq:UncertaintyProt2} \Delta t_{BA} = \omega^{-1}
O\Bigl(\frac{\log N_c}{N_c}\Bigr) \,.
\end{equation}
This result is very remarkable.  Comparing
(\ref{eq:UncertaintyProt2}) and (\ref{eqn:DtSimpleOneWay}) we
observe a near quadratic improvement over the simple protocols
presented above.  This protocol beats the SQL of $1/\sqrt{N_c}$, yet
does not require entangled states or collective measurements.
Without using any of these hallmarks of quantum algorithms, we have
still managed to beat the SQL through the use of an increased
complexity in coherent qubit communications.

\section{Comparison with entanglement}

We now compare the performance of our improved protocol with
alternatives.  The results in~\cite{Bol96} suggest that there should
be a ticking qubit protocol using maximally entangled states that
also beats the SQL.  We briefly present such a protocol, and
demonstrate that our improved protocol gives identical performance.

Consider a protocol similar to our simplest protocol, where the
single ticking qubit is replaced by a $M$-qubit entangled state,
i.e., Alice sends to Bob $M$ qubits in the state,
\begin{equation}
  \label{eq:EntStates}
  \ket{\psi}_A = \tfrac{1}{\sqrt{2}}\bigl[\ket{000..} + \ket{111...}\bigr]_A =
  \tfrac{1}{\sqrt{2}}\bigl[\ket{\mathbf{0}} + \ket{\mathbf{M}}\bigr]_A\,.
\end{equation}
Bob performs $H_B^{\otimes M}$ (i.e., the operation $H_B$ on each qubit) and then
measures the observable $O_B = (-1)^M \sigma_{Z}^{\otimes M}$. The
expectation value of the observable is,
\begin{equation}
  \bra{\psi_B} O_B \ket{\psi_B} = \cos(M \omega t_{BA}) \,,
  \label{eqn:MEntangledO}
\end{equation}
with uncertainty $\Delta O_B = \sin{(M\omega t_{BA})}$.  If $t_{BA}$
is initially known to an accuracy within $\pi/(M\omega)$, then this
procedure gives an estimate of $t_{BA}$ with uncertainty
\begin{equation}
  \Delta t_{BA} = \frac{1}{M \omega}\,.
\end{equation}

A comparison of (\ref{eqn:MEntangledO}) and
(\ref{eqn:ExpectOSimpleOneWay}) reveals that the use of an $M$-qubit
entangled state has produced a single quantum object ticking at an
effective frequency $M \omega$, thus displaying a $1/\sqrt{M}$
performance improvement over using $M$ qubits with frequency
$\omega$ in the simple protocol.  If the number of qubits entangled,
$M$, is allowed to increase with the number of qubits sent, $N$,
then the use of such a scheme can beat the SQL.

Note, however that requirements on the \emph{initial} uncertainty in
$t_{BA}$ are much more stringent.  To give a fair comparison with
our improved protocol, we now present a clock synchronization method
for using entangled states of the form~(\ref{eq:EntStates}) when the
initial uncertainty is comparable to that discussed in the previous
sections, i.e., $\Delta t_{\rm init} \sim \pi/\omega$.  The problem
with the use of an $M$-qubit entangled state as described above is
that it estimates only the least significant digits in $T = \omega
t_{BA}/\pi$, while being unable to estimate the most significant
digits due to its high effective frequency. To estimate $k$ digits
of $T$ with probability of error less than $\epsilon$ using an
entangled protocol, we use the phase estimation algorithm presented
in Sec~\ref{sec:Improved}.  Each of the $k$ bits is estimated
independently.  In the estimation of the $j$th bit, we replace the
$2^j$ coherent exchanges of a single qubit with the single exchange
of a $2^j$-qubit maximally entangled state
$\tfrac{1}{\sqrt{2}}\bigl[\ket{\mathbf{0}} +
\ket{\mathbf{2^j}}\bigr]_A$.  Alice sends this state to Bob, who
performs the operation $H_B^{\otimes 2^j}$ and measures the
observable $O_B = (-1)^{2^j} \sigma_{Z}^{\otimes 2^j}$ to gain an
estimate for the $j$th bit of $T$.  Alice and Bob repeat this enough
times to guarantee a bound on the error probability of
$\varepsilon/k$ for this bit.  It is straightforward to show that
the performance of this algorithm is identical to our improved
protocol, i.e., requiring $O(2^{k} \ln (2k/\epsilon))$ total qubit
communications to estimate $T$ to $k$ bits of precision with total
error probability less than $\epsilon$.  Again, this result
expressed in terms of uncertainty is
\begin{equation}
  \label{eq:EntUncertainty} \Delta t_{BA} = \omega^{-1}
  O\Bigl(\frac{\log N_c}{N_c}\Bigr) \,.
\end{equation}
The additional logarithmic term (when compared with the usual
Heisenberg limit~\cite{Gio04} of $1/(\omega N_c)$) arises from the
need to estimate \emph{all} $k$ digits of $T$, rather than just the
least significant, and to bound the error each digit to at most
$\varepsilon/k$.  Thus, our improved algorithm requiring no
entanglement performs equally well to an algorithm making use of
entanglement.  We conjecture that this scaling ($\Delta t_{BA} =
\omega^{-1} O((\log N_c)/N_c)$) is the fundamental (Heisenberg)
limit for clock synchronization using qubits of frequency $\omega$,
starting with an initial uncertainty of $\Delta t_{\rm init} \sim
\pi/\omega$.

\section{Clock synchronization in the presence of noise}

As discussed in~\cite{Gio01,Gio02}, the performance of an Einstein
synchronization protocol using maximally entangled states
deteriorates with the presence of photon loss.  The same is true of
our protocols, in particular of the improved protocol where multiple
coherent communications are required.  However, we now show that our
improved protocol can still beat the SQL using a lossy channel,
albeit only up to a limited precision determined by the amount of
noise.

Let $\eta$ be the probability that a qubit will not be lost during a
single one-way transmission between Alice and Bob.  The expected
number of runs required to send one qubit from Alice to Bob is:
\begin{equation}
  E(1) = \sum_{n=0}^{\infty} n \eta(1 - \eta)^{n-1} = 1/\eta \,.
\end{equation}

In the simple protocol of Sec.~\ref{sec:Simple}, each qubit
transmission is independent, so the total number of qubit
communications required to achieve a precision of $k$ bits with
probability of error $\leq \epsilon$ is,
\begin{equation}
  N_c = \eta^{-1} \frac{32}{\pi^2} \ln (2 / \epsilon)2^{2k} \,.
\end{equation}

Analysis of the improved protocol of Sec.~\ref{sec:Improved} is more
complex.  Recall the algorithm consisted of $k$ independent rounds.
The $j$th round required $2^j$ coherent communications of a single
qubit, repeated $n = 32 \ln(2k/\varepsilon)$ times.  Because the
$2^j$ communications must be coherent, a loss of a qubit at any
stage will require this step to restart from the beginning.  It will
be convenient to work in ``bounces'' (transfers from Alice to Bob
and then back to Alice). The expected number of bounces to achieve
one successful bounce is $E_B(1) = 1/\eta^2$.  Also, given the
expected number of bounces to achieve $k$ successful bounces,
$E_B(k)$, the expected number required to achieve $k+1$ is $E_B(k+1)
= (E_B(k) + 1)/\eta^2$.  This iterative formula gives,
\begin{equation}
  E_B(k) = \frac{\eta^{-2k}-1}{1-\eta^2} \,.
\end{equation}
Thus the total number of communications required by the improved
algorithm to achieve a precision of $k$ bits with probability of
error $\leq \epsilon$ is:
\begin{equation}
  N_c = 32 \ln (2k / \epsilon) 2\sum_{j=0}^{k-1} E_B(2^j)
  \,.
\end{equation}

\begin{figure}
\begin{center}
\includegraphics[width=3.25in]{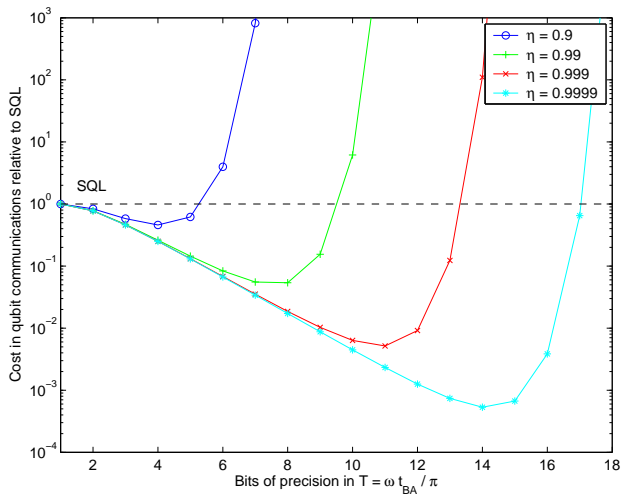}
\end{center}
\caption{Resources requirements of the improved protocol relative to
the simple protocol (SQL) for several values of channel efficiency.}
\label{Fig:ImprovedPInLoss}
\end{figure}

Figure \ref{Fig:ImprovedPInLoss} shows a comparison of the relative
cost (in terms of qubit communications) of the improved protocol
compared with the simple protocol (SQL) with the same noise.  The
improved protocol beats the SQL for low precisions, even in the
presence of small amounts of loss in the channel.  Performance falls
off for high precisions at a threshold dependent on the channel
quality.  The observed behaviour is that the maximum number of bits
of precision scales as $k_{\rm max} \sim \ln(1-\eta)$.  The improved
protocol will beat the SQL for a range of precisions, and this range
increases with channel quality.

Beyond the point where the improved protocol no longer beats the
SQL, one can use a ``hybrid'' algorithm, which uses the improved
protocol to estimate the first $k_1$ bits of precision and then it
switches to a simple protocol to estimate the remaining $k-k_1$
bits. However, in the simple protocol phase, each qubit can be
communicated $2^{k_1}$ times before the measurement is performed
instead of just once, yielding an effective qubit frequency of
$2^{k_1}\omega$.  The performance of such a protocol will scale as
the improved protocol up to $k_1$ bits of precision, and then scale
as the SQL (albeit with a higher effective frequency) once the
simple protocol takes over. Thus the performance relative to the
SQL is constant for precisions of $k_1$ and higher.

As an example, future optical clock standards such as those
described in Takamoto \emph{et al.}~\cite{Tak05} using Sr atoms with
a transition frequency of 429 THz, are predicted to have a
fractional time uncertainty of one part in $10^{18}$. Suppose we
wish to synchonize two such clocks, every second, and thus require
synchronization to $10^{-18}$ seconds. In principle, synchronization
can be achieved via the exchange of Sr atoms (or any ticking qubits
operating at this optical frequency, such as optical photons),
requiring $k=11$ bits of precision in $T$.  For channel transmission
coefficients of 0.9, 0.99, and 0.999, the ``hybrid'' algorithm beats
the SQL by factors of 7, 60 and 400 respectively in qubit
communication cost. Alternatively, exchange of qubits ticking at RF
frequency (such as Cs atoms using the standard 9~192~631~770~Hz
transition) would require $k=26$ bits of precision in $T$.  The
performance relative to the SQL remains approximately the same in
this case, due to the channel noise forbidding operation at the
Heisenberg limit beyond a precision $k_{\rm max} \sim \ln(1-\eta)$.

\section{Discussion}

We have presented an analysis of ``ticking qubit'' protocols for
clock synchronization, based on the Eddington slow clock transport
protocol.  We have demonstrated a simple protocol that through $N_c$
independent one-way qubit communications achieves
 the ``standard quantum limit'' scaling of
$\Delta t_{BA} \sim 1/(\omega \sqrt{N_c})$. This limit can be beaten
by an improved protocol, which requires multiple coherent
communications of a qubit, but makes no use of entanglement and
gives $\Delta t_{BA} \sim (\omega N_c)^{-1}\log N_c$. This result is
in contrast with a standard assumption that entanglement is required
to beat the standard quantum limit.

Inspection of (\ref{Eqn:ExpectOImproved}) reveals that our improved
protocol gives us a way of effectively increasing the qubit ticking
frequency.  By coherently exchanging a qubit back and forth $M$
times, interspersed with each party's $X_P$ operation, we produce an
effective ticking frequency of $M\omega$, precisely like a single
exchange of an $M$-qubit entangled state. Thus, we have the
following three-way equivalence in resources for clock
synchronization in terms of the measurement statistics they
generate:
\begin{enumerate}
\item sending a single qubit ticking at frequency $M\omega$;
\item sending an $M$-qubit entangled state of the form of
Eq.~(\ref{eq:EntStates}), with each qubit at frequency $\omega$;
\item $M$ two-way coherent communications of a single qubit at frequency
$\omega$.
\end{enumerate}
If we consider the number of qubit communications as determining the
cost of a protocol, then we find that techniques (2) and (3) are
equivalent, while (1) gives an improvement by a factor of $M$.
Clearly, there is enormous advantage to possessing qubits with
larger frequencies, as was demonstrated in the second protocol of
Chuang~\cite{Chu00}.  The use of coherent communications as a
substitute for sending multiple-qubit entangled states in quantum
information protocols, first suggested in~\cite{Rud03}, has been
largely unexplored.

Finally, we found that the improved protocol can function in the presence of
noise (a lossy channel).  While the noise limits the maximum precision,
it is still possible to beat the standard quantum limit.

\begin{acknowledgments}
  We thank Andrew Doherty, Terry Rudolph and Robert Spekkens for helpful discussions.
  This project was supported by the Australian Research Council.
\end{acknowledgments}

\end{document}